\documentclass{ws-ijmpa}

\newcommand{\be}{\begin{equation}}
\newcommand{\ee}{\end{equation}}
\newcommand{\bea}{\begin{eqnarray}}
\newcommand{\eea}{\end{eqnarray}}
\newcommand{\bean}{\begin{eqnarray*}}
\newcommand{\eean}{\end{eqnarray*}}
\newcommand{\ba}{\begin{array}}
\newcommand{\ea}{\end{array}}

\newcommand{\slashs}[1]{\not{\!#1}}
\newcommand{\slashm}[1]{\ooalign{\hfil/\hfil\crcr$#1$}}

\begin{document}

\markboth{H.~Kawamura, J.~Kodaira and K.~Tanaka}
{Unraveling Soft Components in the Shape Function 
for Inclusive B Decays}

\catchline{}{}{}{}{}

\title{UNRAVELING SOFT COMPONENTS IN THE\\
SHAPE FUNCTION FOR INCLUSIVE B DECAYS
}

\author{HIROYUKI KAWAMURA}

\address{Radiation Laboratory, RIKEN,
Wako, Saitama 351-0198, JAPAN}

\author{JIRO KODAIRA}

\address{Theory Division, KEK,
Tsukuba, Ibaraki 305-0801, JAPAN}

\author{KAZUHIRO TANAKA}

\address{Dept. of Physics, Juntendo University, Inba-gun, 
Chiba 270-1695, JAPAN}


\maketitle

\begin{abstract}
We present a systematic study of the shape function 
for inclusive $B$-meson decays in the heavy-quark limit,
which is based on the QCD equations of motion and 
heavy-quark symmetry, and takes into account the cusp divergence due to radiative corrections.
%
\end{abstract}


\vspace{0.4cm}
\noindent
Inclusive $B$-meson decays, such as 
semileptonic 
$B \rightarrow X_{u} \ell \bar{\nu}$ decays
and penguin-induced $B \rightarrow X_{s}\gamma$ decays, 
are of special interest because they are sensitive probes
of electroweak parameters as well as new physics.
Especially for the precise determination of the CKM matrix element $|V_{ub}|$,
the region 
near the kinematic endpoint for the lepton energy spectrum, $E_{\ell} \sim E_{\ell}^{\rm max}=m_{B}/2$,
in the decay $B \rightarrow X_{u} \ell \bar{\nu}$ 
plays an important role,\cite{Kakuno:2003fk}
avoiding large backgrounds from the decays into charmed particles,
$B \rightarrow X_{c} \ell \bar{\nu}$.
In this endpoint region, the hadronic decay products 
evolving from the $u$-quark 
have large energy but small invariant mass, so that
the hadronic final state becomes jet-like with collinear interactions of an outgoing light quark.
This implies that 
the corresponding differential decay rates are expressed as
the light-cone expansion, 
analogously with
deep inelastic lepton-nucleon scattering (DIS) cross sections:
the leading term in the light-cone expansion
is described by an analogue of the leading twist in the DIS, 
i.e., by a factorization formula
where 
a structure function with
all nonperturbative, 
long-distance ($\sim 1/\Lambda_{\rm QCD}$) contribution
is convoluted with the perturbatively calculable 
function.\cite{Korchemsky:1994jb}
%
This
factorization formula 
is valid in the leading power of $\Lambda_{\rm QCD}/m_{b}$
and to all orders in $\alpha_{s}$,\cite{Korchemsky:1994jb,Bauer:2003pi,blnp} 
and the structure function, called the shape function,
is expressed
as 
the $B$-meson
matrix element of a 
bilocal light-cone operator 
in the heavy-quark effective theory (HQET):\cite{Korchemsky:1994jb}
\begin{equation}
\langle \bar{B}(v) | \bar{h}_{v}(tn) 
{\rm P} 
e^{ig\!\! \int_0^t \!\! d\xi\, n\cdot A (\xi n)}
h_{v}(0) |\bar{B}(v) \rangle
= \tilde{f}(t)= \int d\omega e^{i\omega t} f(\omega) \ ,
\label{eq:sf0}
\end{equation}
where $v^{\mu}$ is the 4-velocity of the $B$-meson ($v^2 =1$), and
$n^{\mu}$ denotes a light-like vector ($n^{2}=0,\ v\cdot n=1$), 
pointing in the direction of the final-state light-quark jet.
%
$h_{v}(x)$ denotes the effective $b$-quark field,
$b(x) \approx \exp(-im_{b} v\cdot x)h_{v}(x)$,
satisfying
$\slashm{v} h_{v} = h_{v}$.\cite{Neubert:1994mb}
$\tilde{f}(t)$ and $f(\omega)$ denote the shape functions in the coordinate
and (residual) momentum representations, 
respectively.
The shape function
is process independent 
so that
(\ref{eq:sf0}) describes 
other inclusive $B$ decays,
e.g., the photon energy spectrum 
of
$B \rightarrow X_{s}\gamma$.
For brevity, we do not show
the path-ordered gauge factors 
connecting the constituent fields in the following.

%
The bilocal operator of
(\ref{eq:sf0}) 
is renormalized at the scale $\mu$.
Recently it has been emphasized that 
the radiative corrections 
induce a Sudakov-type
strong scale dependence on (\ref{eq:sf0}),
governed by the 
``cusp anomalous dimension'',\cite{Korchemsky:1994jb,Bauer:2003pi,blnp}
and, as a result,
$\tilde{f}(t)$ 
is singular for short distances, 
$\tilde{f}(t) \sim -(\alpha_s C_F /\pi )\log^2 (it\mu)$ as $t \rightarrow 0$; 
for $f(\omega)$,
this implies that ``radiative tail'' is generated for $\omega\ll -\mu$, and 
all non-negative moments $\int d\omega \omega^j f(\omega)$
become UV divergent.\cite{Korchemsky:1994jb,Bauer:2003pi,blnp}
This contamination of the short-distance singularity, that cannot be removed by the
usual renormalization, suggests\cite{Korchemsky:1994jb,Bauer:2003pi} 
that the bilocal operator of (\ref{eq:sf0})
has to be further decomposed as
\be
\bar{h}_{v}(tn) 
h_{v}(0) =\sum_{\cal O} C_{\cal O}(t, \mu){\cal O} (\mu)
\label{OPE}
\ee
with the Wilson coefficient functions $C_{\cal O}(t, \mu)$ that represent the ``hard components''
in (\ref{eq:sf0}), integrating out the modes associated with the mass scales $\gtrsim 1/t$; they 
are perturbatively calculable
as $C_{\cal O}=C_{\cal O}^{(0)}+(\alpha_s /\pi )C_{\cal O}^{(1)} + \cdots$, 
and separate out all short-distance 
singularities of (\ref{eq:sf0}), 
$C_{\cal O}^{(1)}(t, \mu) \sim C_F \log^2 (it\mu)$ as $t\rightarrow 0$. 
On the other hand, 
the local operators
${\cal O}(\mu)$ represent the ``soft components'' 
and contain all nonperturbative effects.
The OPE (\ref{OPE}) is useful for $t \mu \lesssim 1$, and 
an independent set of 
${\cal O}$ and the corresponding 
coefficients $C_{\cal O}$ 
can be determined by matching both sides of (\ref{OPE})
at the scale $\mu \simeq 1/t$.
%
We perform the matching, order by order in $\alpha_s$, 
and derive  
the sophisticated operator structure of (\ref{OPE}),\cite{KKT}
that was previously unknown.

$C_{\cal O}^{(0)}$ at LO
are independent of 
$\mu$; therefore,
$C_{\cal O}^{(0)}$ does not have any noncanonical 
$t$-dependence 
singular for $t \rightarrow 0$.
Thus, in the tree-level matching of (\ref{OPE}), 
$\bar{h}_{v}(tn)h_{v}(0)$ can be Taylor expanded about $t=0$,
and one can 
treat the corresponding local operators 
$\bar{h}_{v}D_{\mu_1}\cdots D_{\mu_j} h_{v}$.
However, this actually requires a complicated task to disentangle 
$\bar{h}_{v}D_{\mu_1}\cdots D_{\mu_j} h_{v}$ 
by a basis of ``canonical operators'' using 
the HQET equations of motion (EOM), $iv\cdot D h_v =0$,
and has been feasible for the first few 
operators.\cite{Korchemsky:1994jb}
Here we solve the EOM constraints 
for the 
nonlocal operators, and
Taylor expand the results at the final step:\cite{KKT}
in the exact operator identity,
\be 
v^{\mu}\frac{\partial}{\partial x^{\mu}}\bar{h}_{v}(x) 
h_{v}(0) = 
\bar{h}_{v}(x)
v\ \cdot\! 
\stackrel{\leftarrow}{D} h_{v}(0)
+ i \int_{0}^{1}duu \ \bar{h}_{v}(x) gG_{\mu \nu}(ux) v^{\mu} x^{\nu}
  h_{v}(0) 
\label{eq:id}
\ee
with $G_{\mu \nu}= (i/g)[D_{\mu}, D_{\nu}]$,
the first term in the RHS vanishes using the EOM.
Because $x^{\mu}$ is not restricted on the light cone,
the corresponding constraint equation for $x_{\mu}\rightarrow tn_{\mu}$ is derived by
combining with the light-cone expansion,
%
\be
\bar{h}_{v}(x) h_{v}(0) = \left[ \bar{h}_{v}(x) h_{v}(0) \right]_{\rm lt}
+ \frac{x^{2}}{4} \int_{0}^{1}\! \frac{du}{u}\
\frac{\partial^{2}}{\partial x_{\mu} \partial x^{\mu}}
\bar{h}_{v}(ux) h_{v}(0)+ {\cal O}(x^{4})\ ,
\label{eq:formal2}
\ee
where all local operators arising in the Taylor expansion of the first term 
are traceless and the following terms represent the corresponding ``trace part''. We get
\bea
t\frac{d}{dt}&& \!\!\! \bar{h}_{v}(tn) h_{v}(0) 
+\bar{h}_{v}(tn) h_{v}(0) -\bar{h}_{v}(0) h_{v}(0)= J(t)\ ,
\label{eq:d}\\
J(t)&& \!\!\! = t^2 \!\! \int_0^1 \!\!\! du  \! 
\left( \! 2iu\bar{h}_v (tn)gG_{\mu \nu } (utn)v^\mu  n^\nu  h_v (0) 
- \left. {\frac{\partial^{2}}{u \partial x_{\mu} \partial x^{\mu}}
  \bar{h}_v (ux)h_v (0)} \right|_{x \to tn}  \right).
\label{eq:j}
\eea
Here the second and third terms in the LHS of (\ref{eq:d}) 
give the analogue of Nachtmann's correction in the DIS,
due to deviation from the light-cone in $\left[ \bar{h}_{v}(x) h_{v}(0) \right]_{\rm lt}$ 
of (\ref{eq:formal2}). 
%
Disentangling the last term of (\ref{eq:j}) with another operator identity 
from the EOM for heavy-quark and gluon fields,\cite{KKT} the ``source'' term (\ref{eq:j})
becomes ($D_{\perp \mu}
\equiv D_{\mu}-v_{\mu}v \cdot D$)
\bea
\!\! J(t) &&\!\!\!
=  \! - \! \int_0^t \! \! d\tau \! \left( {\tau \bar{h}_v (\tau n)
\overleftarrow{D}_{\perp}^2  
h_v (0)} \right. 
\! +\!  i\tau ^2 \!\! \int_0^1 \!\!\! duu^2 
\bar{h}_v (\tau n)g^2 t^a \bar{q}(u\tau n)t^a \!\! \slashs{n} q(u\tau n)h_v (0)  
\nonumber\\ 
&&+ \left.  2\tau ^3 \!\! \int_0^1 \!\!\! duu \int_0^u \!\!\! dss 
\bar{h}_v (\tau n)gG_{\mu \nu } (u\tau n)n^\nu  
gG^{\mu \rho } (s\tau n)n_\rho  h_v (0)\! \right)\ ,
\label{eq:j2}
\eea
where the first term is nonlocal version of the kinetic energy operator.
Eqs.~(\ref{eq:d}), (\ref{eq:j2}) show that the $t$-dependence, i.e., the longitudinal-momentum dependence,
of the shape function is controlled by higher dimensional operators, 
that represent Fermi motion effects and 4-particle correlations with additional 
quarks and gluons inside the $B$ meson, and 
are immediately integrated and Taylor expanded about $t=0$: 
\bea
 \bar h_v (tn)h_v (0) &&= \bar h_v (0)h_v (0) + \frac{1}{t}\int_0^t {d\tau J(\tau )}  
\nonumber\\ 
 && = \bar h_v h_v  - \frac{t^2}{6}\bar h_v D_{\perp}^2 h_v  
- \frac{it^3}{36}\bar h_v g^2 t^a \bar qt^a \!\! \slashs{n} q h_v 
+ 
\left. \frac{t^4}{120}\right( \bar h_v  D_{\perp}^4  h_v  
\nonumber\\
&&\left.  - \frac{2i}{3}\bar h_v g^2 t^a \bar q t^a \gamma_\mu  q D_{\perp}^{\mu} h_v 
 - \frac{3}{2}\bar h_v gG_{\mu \nu } n^\nu  gG^{\mu \rho } n_\rho  h_v  \right) +  \cdots\ ,
\label{eq:tree} 
\eea
where we have simplified the tensor structure of the operators in the RHS
using Lorentz invariance; for the $j$-th power $t^j$, the local operators of dimension $j+3$ contribute.
The expression for the general $j$-th power shows\cite{KKT} that
an increasing number of operators participate for increasing $j$
(i.e., $\sim j^2/4$ operators for $j \gg 1$). The result (\ref{eq:tree}) gives (\ref{OPE})
with an operator basis 
$\{{\cal O}\}=\{\bar h_v h_v , \bar h_v D_{\perp}^2 h_v , 
\bar h_v g^2 t^a \bar qt^a \slashm{n} q h_v , \cdots\}$ 
and the corresponding LO
coefficients $C_{\cal O}^{(0)}$.

Utilizing this operator basis,
we proceed to the one-loop matching of (\ref{OPE})
to obtain 
$C_{\cal O}^{(1)}$. For the bilinear-type operators 
${\cal O}=\bar h_v h_v$ and $\bar h_v D_{\perp}^2 h_v$, 
the calculation is straightforward. We consider the two-point heavy-quark Green function
with the insertion of (\ref{OPE}), and compute the one-loop corrections
for the nonlocal and local operators corresponding to LHS and RHS of (\ref{OPE}), 
respectively: for $\bar{h}_v (tn)h_v (0)$ this was calculated 
in the coordinate representation in Ref.\cite{Korchemsky:1994jb};
the number density operator $\bar h_v h_v$ and the kinetic energy operator $\bar h_v D_{\perp}^2 h_v$ 
obey non-renormalization.\cite{Neubert:1994mb}
We get
$C_{\bar{h}_v h_v}^{(1)}(t, \mu)= - C_F 
[ \log^2 (it\mu e^{\gamma_{E}}) - \log(it\mu e^{\gamma_{E}}) + 5\pi ^2 /24 ]$,
$C_{\bar{h}_v D_{\perp}^2 h_v}^{(1)}(t, \mu)=-t^2 C_{\bar{h}_v h_v}^{(1)}(t, \mu)/6$
%
%
in the $\overline{\rm MS}$ scheme.
Going over to the $\omega$-representation via (\ref{eq:sf0}),
this result coincides with the corresponding coefficients 
obtained in Ref.\cite{Bauer:2003pi}, and is consistent with those calculated in the
momentum cutoff scheme.\cite{blnp}
Matching for the 
other operators, 
${\cal O}= \bar h_v g^2 t^a \bar qt^a  \slashm{n} q h_v , \cdots$,
can be treated similarly, but the complications arise  
because of multiparticle nature of the relevant operators.

Substituting all above results into (\ref{OPE}), we get the OPE for the shape function
valid up to $O\left(\alpha_s^2 \right)$ 
and $O\left((\Lambda_{\rm QCD} t)^3 \alpha_s \right)$ corrections: for 
low dimensional operators ${\cal O}=\bar h_v h_v , \bar h_v D_{\perp}^2 h_v$, the coefficients $C_{\cal O}$ are 
obtained  
at the NLO accuracy, 
and matrix elements of those operators 
directly 
give 
the familiar HQET parameters\cite{Neubert:1994mb}
at the scale $\sim$1GeV, as $\langle \bar{B}(v) | \bar h_v h_v |\bar{B}(v) \rangle =1$,
$\langle \bar{B}(v) | \bar h_v D_{\perp}^2 h_v|\bar{B}(v) \rangle =- \lambda_1$. For 
higher dimensional operators ${\cal O}= \bar h_v g^2 t^a \bar qt^a  \slashm{n} q h_v$,
$\bar h_v  D_{\perp}^4  h_v$,
$\bar h_v g^2 t^a \bar q t^a \gamma_\mu  q D_{\perp}^{\mu} h_v$,
$\bar h_v gG_{\mu \nu } n^\nu  gG^{\mu \rho } n_\rho  h_v  , \cdots$, 
the coefficients $C_{\cal O}$ are 
obtained at the LO, and matrix elements of those operators are related to 
the nonperturbative parameters
at 1GeV scale through evolution governed by the corresponding anomalous dimension. 
There are some simple estimates of those nonperturbative parameters.\cite{KKT}
The anomalous dimension
of $\bar h_v g^2 t^a \bar qt^a  \slashm{n} q h_v$ is known at one-loop,\cite{Bauer:2003pi} 
while the anomalous dimension matrix describing the renormalization mixing among
$\{\bar h_v  D_{\perp}^4  h_v ,\
\bar h_v g^2 t^a \bar q t^a \gamma_\mu  q D_{\perp}^{\mu} h_v ,\
\bar h_v gG_{\mu \nu } n^\nu  gG^{\mu \rho } n_\rho  h_v \}$ is not known.
We note that matrix elements of the $j+3$ dimensional
operators generate $O\left((\Lambda_{\rm QCD}t)^j \right)$
power corrections in the OPE (\ref{OPE}) because canonical size of 
the corresponding nonperturbative parameters is $\sim \Lambda_{\rm QCD}^j$;
for the regions with $m_{b} \gg 1/t$,
these power corrections are {\it enhanced} compared with the usual subleading corrections,
which are suppressed by powers of $\Lambda_{\rm QCD}/m_{b}$.
Physically, this corresponds to the situation in which the decay spectra in the endpoint region 
is smeared over a range $\Delta$ with $m_{b} \gg \Delta \gtrsim \Lambda_{\rm QCD}$.\cite{Bauer:2003pi}
When $t \simeq 1/\Lambda_{\rm QCD}$, corresponding to $\Delta \simeq \Lambda_{\rm QCD}$, 
the higher-order power corrections in (\ref{OPE})
are not suppressed, and the OPE has to be resummed. Under certain assumptions, 
our differential equation (\ref{eq:d}) gives a ``resummed'' solution,\cite{KKT}
whose behavior is consistent with a model\cite{Korchemsky:1994jb} 
guided by IR renormalon analysis.

In conclusion, our results for (\ref{OPE}) show that
the shape function is a much more complicated object than the simple momentum distribution
of the $b$-quark inside the $B$-meson,
due to novel behavior of perturbative as well as nonperturbative nature.

\vspace{-0.3cm}

\section*{Acknowledgements}
The work of J.K. and K.T. was supported by the Grant-in-Aid for 
Scientific Research Nos. C-16540255
and C-16540266. 


\end{document}